\documentclass[sigconf]{acmart}
\AtBeginDocument{%
  }

\setcopyright{acmlicensed}
\copyrightyear{2026}
\acmYear{2026}
\acmDOI{}
\acmConference[CUG'26]{Cray User Group}{April 26--30,
  2026}{Nice, France}
\acmBooktitle{conditionally accepted for publication in the ACM International Conference Proceedings Series (ICPS).}

\acmISBN{}

\usepackage[acronym]{glossaries}
\usepackage{xcolor}
\usepackage{listings}
\usepackage{array}
\usepackage{tikz}
\usetikzlibrary{positioning,fit,arrows.meta}

\begin{document}

\title{Enabling Hybrid HPCQC Workflows with a Heterogeneous Software Stack}


\author{Muhammad Nufail Farooqi}
\affiliation{%
  \institution{Leibniz Supercomputing Centre (LRZ)}
  \city{Garching}
  \country{Germany}}
\email{muhammad.farooqi@lrz.de}

\author{Minh Chung}
\affiliation{%
  \institution{Leibniz Supercomputing Centre (LRZ)}
  \city{Garching}
  \country{Germany}}
\email{minh.chung@lrz.de}

\author{Burak Mete}
\affiliation{%
  \institution{Leibniz Supercomputing Centre (LRZ)}
  \city{Garching}
  \country{Germany}}
\email{burak.mete@lrz.de}

\author{Eric Mansfield}
\affiliation{%
  \institution{IQM Quantum Computers (IQM)}
  \city{Munich}
  \country{Germany}}
\email{eric.mansfield@meetiqm.com}

\author{Bernd Hoffmann}
\affiliation{%
  \institution{IQM Quantum Computers (IQM)}
  \city{Munich}
  \country{Germany}}
\email{bernd.hoffmann@meetiqm.com}

\author{Teemu Mattsson}
\affiliation{%
  \institution{IQM Quantum Computers (IQM)}
  \city{Munich}
  \country{Germany}}
\email{teemu.mattsson@meetiqm.com}

\author{Laura Schulz}
\affiliation{%
  \institution{Argonne National Laboratory (ANL)}
  \city{Lemont}
  \state{Illinois}
  \country{USA}}
\email{schulz@anl.gov}

\author{Jorge Echavarria}
\affiliation{%
  \institution{Munich Quantum Valley (MQV)}
  \city{Garching}
  \country{Germany}}
\email{jorge.echavarria@munich-quantum-valley.de}


\renewcommand{\shortauthors}{Farooqi et al.}

\newacronym{api}{API}{\textit{Application Programming Interface}}
\newacronym{cpu}{CPU}{\textit{Central Processing Unit}}
\newacronym{eqs3}{EQS3}{\textit{European Quantum Systems and Software Summit}}
\newacronym{fpga}{FPGA}{\textit{Field-Programmable Gate Array}}
\newacronym{gpu}{GPU}{\textit{Graphics Processing Unit}}
\newacronym[\glslongpluralkey={\textit{Generic RESources}}, \glsshortpluralkey={GRES}]{gres}{GRES}{\textit{Generic RESource}}
\newacronym{grape}{GRAPE}{\textit{Gradient Ascent Pulse Engineering}}
\newacronym{hpc}{HPC}{\textit{High Performance Computing}}
\newacronym{hpcqc}{HPCQC}{\textit{High Performance Computing-Quantum Computing}}
\newacronym{hpe}{HPE}{\textit{Hewlett Packard Enterprise}}
\newacronym{ir}{IR}{\textit{Intermediate Representation}}
\newacronym{iqm}{IQM}{\textit{IQM Quantum Computers}}
\newacronym{isv}{ISV}{\textit{Independent Software Vendor}}
\newacronym{lrz}{LRZ}{\textit{Leibniz Supercomputing Centre}}
\newacronym{mlir}{MLIR}{\textit{Multi-Level Intermediate Representation}}
\newacronym{mqss}{MQSS}{\textit{Munich Quantum Software Stack}}
\newacronym{mqv}{MQV}{\textit{Munich Quantum Valley}}
\newacronym{nisq}{NISQ}{\textit{Noisy Intermediate-Scale Quantum}}
\newacronym{ornl}{ORNL}{\textit{Oak Ridge National Laboratory}}
\newacronym{qc}{QC}{\textit{Quantum Computing}}
\newacronym{qdessi}{Q-DESSI}{\textit{Quantum Development Environment, System Software \& Integration}}
\newacronym{qec}{QEC}{\textit{Quantum Error Correction}}
\newacronym{qdmi}{QDMI}{\textit{Quantum Device Management Interface}}
\newacronym{qir}{QIR}{\textit{Quantum Intermediate Representation}}
\newacronym{qis}{QIS}{\textit{Quantum Instruction Set}}
\newacronym{qpi}{QPI}{\textit{Quantum Programming Interface}}
\newacronym{qpu}{QPU}{\textit{Quantum Processing Unit}}
\newacronym{qrm}{QRM\&CI}{\textit{Quantum Resource Manager \& Compiler Infrastructure}}
\newacronym{qrmi}{QRMI}{\textit{Quantum Resource Management Interface}}
\newacronym{qnn}{QNN}{\textit{Quantum Neural Network}}
\newacronym{tem}{TEM}{\textit{Technical Exchange Meeting}}
\newacronym{tum}{TUM}{\textit{Technical University of Munich}}
\newacronym{vqe}{VQE}{\textit{Variational Quantum Eigensolver}}

\newcommand{\je}[1]{{\textcolor{red}{#1}}}

\begin{abstract}


In this work, we demonstrate hybrid \gls{hpcqc} workflows on a production peta\-scale system.
The demonstration combines three components: the SuperMUC-NG supercomputer at the \gls{lrz}, a 20-qubit superconducting quantum processor provided by \gls{iqm}, and \gls{mqv}'s \gls{mqss}.

Integrating quantum processors into \gls{hpc} systems requires a heterogeneous software stack capable of orchestrating classical and quantum resources within established supercomputing workflows.
\gls{mqss} treats \glspl{qpu} as scheduler-managed accelerators and it performs resource coordination following a two-level scheduling scheme.
Slurm performs system-level allocation by exposing \glspl{qpu} as \glspl{gres}, while the \gls{mqss} \gls{qrm} performs just-in-time compilation and subsequent dispatch of quantum circuits.

To integrate with existing \gls{hpc} operations without modifying the scheduler core, \gls{mqss} introduces an open-source \textit{SLURM Plugin Suite} based on Prolog/Epilog scripts and SPANK modules. Experimental results show that hybrid \gls{hpcqc} workflows can be executed without significant latency overhead compared to conventional workloads.

The presented architecture provides a portable integration model for quantum accelerators on large-scale \gls{hpc} systems and is directly applicable to next-generation \gls{hpe} Cray platforms, including \gls{lrz}'s upcoming "Blue Lion" supercomputer.
\end{abstract}


  


\maketitle

\glsresetall

\section{Introduction}
The emergence of quantum computing offers new opportunities to accelerate a range of scientific applications that are traditionally executed on \gls{hpc} systems. One practical way of integrating these technologies is to treat \glspl{qpu} as specialised accelerators within \gls{hpc} infrastructures, in a similar way to \glspl{gpu} and other hardware accelerators. This model enables quantum devices to be incorporated into existing computing environments and leverages the well-established \gls{hpc} ecosystem for resource management and scheduling. It also allows for the development of hybrid \gls{hpcqc} workflows, in which classical and quantum resources collaborate within the same application.

However, integrating \glspl{qpu} into established supercomputing systems presents several challenges. Quantum devices differ significantly from classical accelerators in terms of programming models and execution paradigms, and they are also scarce resources. At the same time, \gls{hpc} platforms rely on stable and scalable scheduling frameworks and operational policies that cannot easily be modified. Addressing these constraints requires a heterogeneous software stack that can integrate quantum hardware into existing \gls{hpc} ecosystems while preserving established workflows, scheduling mechanisms and system-level abstractions.

We describe the design, implementation, and deployment of a heterogeneous software and scheduling stack that enables tightly coupled hybrid \gls{hpcqc} workflows. Our platform integrates the \gls{mqss}, as first introduced in \cite{Burgholzer26}, with \gls{lrz}'s SuperMUC-NG petascale system, and a 20-qubit superconducting quantum processor developed and delivered by \gls{iqm} as part of the Q-Exa project \cite{lrziqm24}. 
\gls{mqss}, developed collaboratively by the \gls{mqv} initiative, \gls{lrz}, and the \gls{tum}, provides a technology-agnostic, layered runtime that exposes quantum processors as scheduler-managed accelerators and presents a unified programming model for hybrid applications \cite{mqss25}. 

In this paper, we will focus on the system-level engineering that allows SLURM~\cite{slurm} to treat quantum processors as first-class resources, on the \textit{MQSS SLURM Plugin Suite} \cite{mqss_slurm_plugins_suite} that implements that integration, and on practical lessons from the SuperMUC-NG + \gls{iqm} deployment \cite{Mansfield25}. We report that hybrid jobs execute with no significant latency overhead relative to comparable purely classical \gls{hpc} jobs, and we discuss how the same stack will operate on \gls{lrz}'s forthcoming \gls{hpe} Cray "Blue Lion" system \cite{lrzhpe24}.
\section{System overview and primary contributions}

\begin{figure*}[htb]
  \centering
  \includegraphics[width=\textwidth]{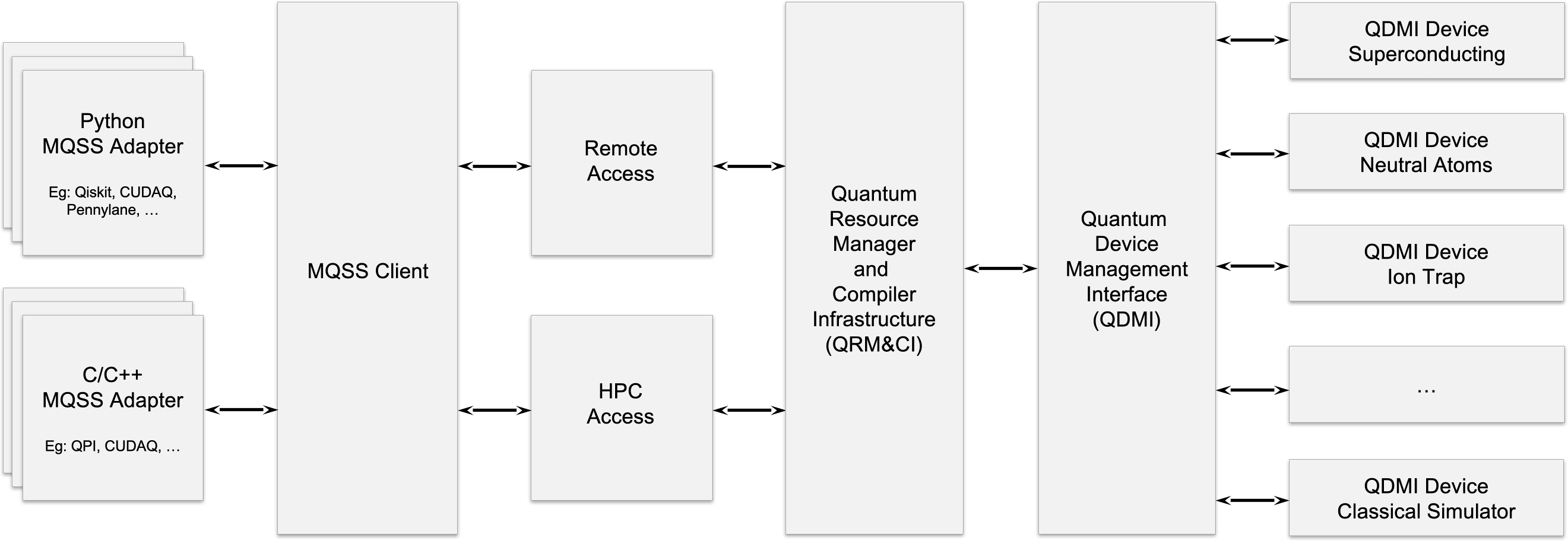}
  \caption{MQSS architecture showing Front-end adapters, QRM\&CI, and QDMI-connected quantum devices.}
  \label{fig:mqss}
\end{figure*}

\gls{mqss} is a modular software stack that bridges high-level hybrid applications and device-specific control. As shown in Fig.~\ref{fig:mqss}, \gls{mqss} includes (1) high-level front-ends and adapters for common quantum programming frameworks~\cite{qiskit2024, pennylane, cudaq, qpi}, (2) a just-in-time compilation layer inside the \gls{qrm}~\cite{Burgholzer26}, and (3) an interface to the hardware, namely the \gls{qdmi}~\cite{qdmi}. The front-end layer provides adapters to widely used quantum programming frameworks and exposes a unified client interface that enables users to submit quantum kernels from different programming environments while remaining agnostic to the underlying hardware. These adapters translate framework-specific representations into \gls{mqss}-compatible intermediate formats and forward them to the runtime system for execution.

The core runtime component of \gls{mqss} is the \gls{qrm}, which combines dynamic compilation, scheduling, and resource management to orchestrate the execution of quantum kernels. It receives circuits in high-level \glspl{ir} and progressively lowers them through a sequence of compilation passes into hardware-specific instructions ready for execution. The compilation pipeline relies on a family of \glspl{ir} and leverages the \gls{mlir} framework to enable modular and extensible compiler pipelines that support both classical and quantum abstractions. This design allows the compiler to perform hardware-aware optimizations, select appropriate devices, and transpile circuits into native gate sets supported by the target hardware.

At the hardware interface layer, \gls{qdmi} provides a standardized communication interface between \gls{mqss} and connected quantum devices. Through this interface, the stack can query real-time device properties such as topology, gate fidelities, calibration status, and system availability, allowing the runtime and compiler infrastructure to adapt execution strategies to the current state of the hardware. \gls{qdmi} also manages job submission, result retrieval, and session management for quantum devices, enabling seamless integration of heterogeneous quantum platforms within the same software stack and facilitating their deployment in \gls{hpc} environments.

The primary contributions of this work are: (1) a practical architecture and two-level scheduling model that integrates SLURM and \gls{mqss}, enabling quantum processors, such as the Q-Exa demonstrator, to be treated as scheduler-managed \gls{gres} accelerators. This model allows \gls{hpc} jobs to request quantum resources alongside classical nodes, while \gls{mqss} handles the device-specific compilation, dispatch, and runtime management of quantum circuits, ensuring seamless execution within existing \gls{hpc} workflows. (2) The \textit{MQSS SLURM Plugin Suite}, a production-grade implementation using SLURM Prolog/Epilog hooks and SPANK modules \cite{mqss_slurm_plugins_suite}, a collection of plugins to interface directly with \gls{mqss} to query real-time device status, manage job submission, and synchronize the allocation of \glspl{qpu} with SLURM's global scheduler, all without requiring intrusive modifications to the core scheduler. (3) An operational case study of the integrated system deployed at \gls{lrz}, linking SuperMUC-NG and \gls{iqm}'s Q-Exa system, which demonstrates the feasibility of running hybrid \gls{hpcqc} workflows on a production-scale supercomputer with minimal latency overhead and practical orchestration of both classical and quantum resources.

\section{Two-level scheduling model} 

Hybrid \gls{hpcqc} workflows require careful coordination between classical and quantum tasks to ensure efficient execution across heterogeneous resources. 
To achieve this, we adopt a two-level scheduling model that separates global resource allocation from device-level quantum execution. 
At the first level, SLURM continues to act as the authoritative scheduler for the \gls{hpc} environment, handling the allocation of classical resources, such as compute nodes and traditional accelerators like \glspl{gpu}. Similar to Viviani et al. ~\cite{paolo2025},
\glspl{qpu} are incorporated into this framework as \gls{gres}, allowing them to be requested in job scripts using parameters such as \verb|--gres=qpu:1|. 
Each compute node may host a single \gls{qpu} by default, but this value can be adjusted to enable sharing of the \gls{qpu} among multiple jobs and achieve the desired level of quantum resource occupancy. 
When a hybrid job is submitted, SLURM ensures that both classical and quantum resources are reserved according to user specifications and cluster policies, providing a unified interface for resource management and enforcing global scheduling constraints while maintaining fairness and efficient utilization of the \gls{hpc} system.

At the second level, the \gls{qrm} component takes over when quantum tasks are offloaded to the \gls{qpu} during execution of the hybrid applications. 
\gls{qrm} handles queuing, just-in-time compilation, hardware-aware optimization, and the dispatch of quantum kernels to the device, ensuring that execution is aligned with the current state and calibration of the processor. 
This design separates responsibilities cleanly: SLURM focuses on cluster-level allocation and scheduling policies, while \gls{qrm} manages device-level sequencing, vendor-specific instructions, and runtime optimizations. 
\gls{qdmi} serves as the critical link between these two levels, providing SLURM and the workflow system with real-time information on \gls{qpu} availability, calibration status, and error rates. By exposing these hardware metrics, \gls{qdmi} enables intelligent scheduling decisions and allows hybrid workflows to execute efficiently across both classical and quantum resources, maintaining high utilization without introducing significant latency overhead.

\subsection{Scheduling policies and predictable execution} 
\gls{mqss} implements the concept of \textit{quantum availability}, a metric that captures current \gls{qpu} load and scheduled calibration overhead. 
\textit{Quantum availability} is supplied to the scheduler logic so that SLURM can avoid starting a job that will immediately stall waiting for the \gls{qpu}. 
We exploit SLURM's backfilling capabilities to maintain overall system utilization: while a quantum-accelerated job is pending due to \gls{qpu} unavailability, SLURM may run other work that does not require the \gls{qpu}. When the \gls{qpu} becomes available, pending quantum jobs are dispatched. This policy yields predictable quantum execution times for users while minimizing idle time for the \gls{qpu}.
\section{MQSS SLURM Plugin Suite} 

\begin{figure*}[t]
  \centering
  \includegraphics[width=\textwidth]{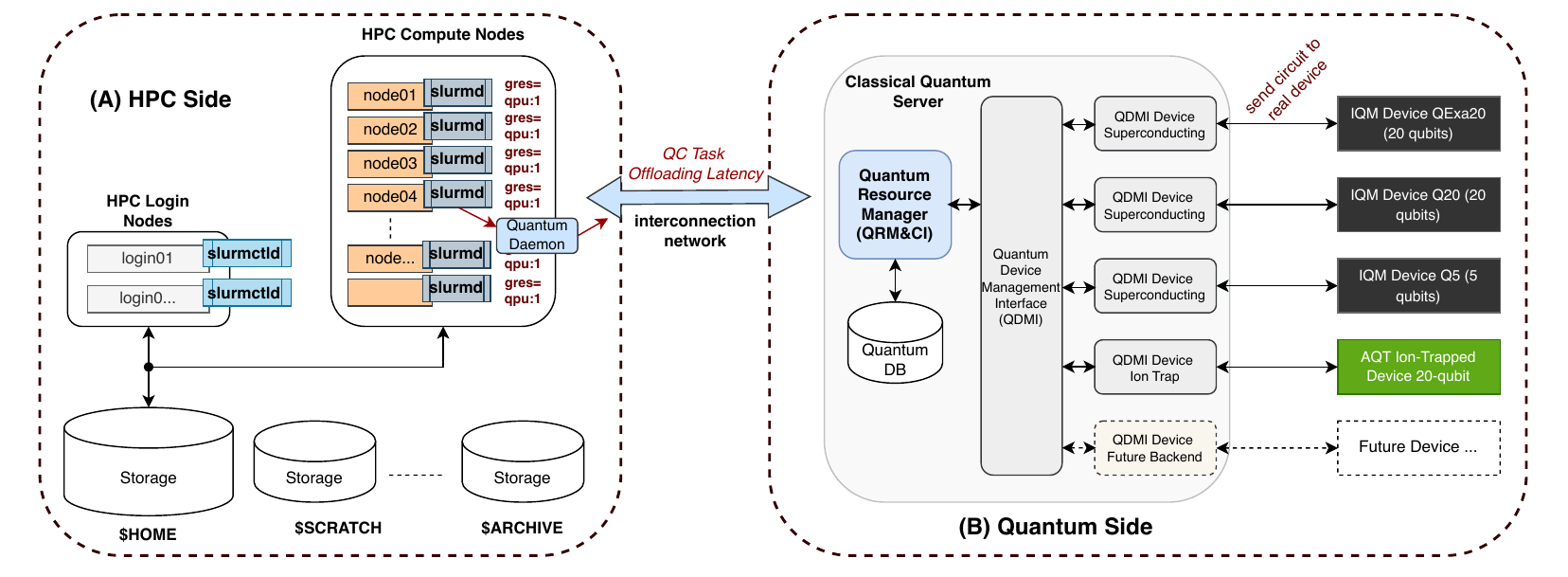}
  \caption{An illustration of HPCQC integration with MQSS, in which (A) shows the HPC side, and (B) shows the quantum side.}
  \label{fig:hpcqc-integration-mqss}
\end{figure*}

To facilitate \gls{hpcqc} job submission, we developed the \textit{MQSS SLURM Plugin Suite}, which was integrated into the production system during the initial deployment phase. While the suite remains under active development, it currently leverages standard SLURM extension mechanisms, specifically Prolog/Epilog scripts and the SLURM Plug-in Architecture for Node and job Control (SPANK) plugins
\cite{slurm_spank}. As shown in Figure~\ref{fig:hpcqc-integration-mqss}, (A) illustrates the login, compute nodes, and components on the \gls{hpc} side, while (B) shows the quantum side with the quantum server deploying \gls{qrm}. 
%
%
\gls{qrm} interfaces with real quantum backends through \gls{qdmi}. It establishes direct connections to specific backends via their respective \textit{QDMI Device} implementations provided by the hardware vendors.

The \textit{MQSS SLURM Plugin Suite} is designed with two primary objectives: to minimize modifications to the core SLURM codebase and to provide a flexible interface for coordinating between classical \gls{hpc} schedulers and quantum-specific schedulers. By decoupling these environments, the suite allows for high-level synchronization between \gls{hpc} and quantum computing.

Technically, Prolog/Epilog scripts or SPANK plugins are SLURM's native extensibility. These mechanisms allow the plugin suite to be deployed, updated, or removed as an external layer, requiring no changes to the SLURM source code. SLURM deployed on production systems launches \texttt{slurmctl} (known as the SLURM controller) and \texttt{slurmd} (SLURM daemon).

\begin{table*}[ht]
    \centering
    \caption{SLURM events for Prologs and Epilogs are available to control Job Allocation, Execution, and Termination, and can also be implemented as SPANK plugins~\cite{jette2023slurmarch}.}
    \label{tab:slurm-phases}

    \begin{tabular}{%
        >{\raggedright}p{0.15\textwidth}%
        >{\raggedright}p{0.17\textwidth}%
        >{\raggedright}p{0.11\textwidth}%
        >{\raggedright}p{0.17\textwidth}%
        p{0.3\textwidth}}
    \toprule
    \textbf{Parameter} & %
    \textbf{Location} & %
    \textbf{Invoked by} & %
    \textbf{User} & %
    \textbf{When executed} \\
    \midrule
    Prolog (specified in \texttt{slurm.conf}) & %
    Compute Nodes & %
    \texttt{slurmd} & %
    SlurmdUser (normally root user) & %
    First job or job step initiation on that node (by default); \texttt{PrologFlags=Alloc} will force the script to be executed at job allocation \\
    \midrule
    PrologSlurmctld (specified in \texttt{slurm.conf}) & Head Nodes or Master Nodes (where \texttt{slurmctld} runs) & \texttt{slurmctld} & SlurmctldUser & At job allocation \\
    \midrule
    Epilog (specified in \texttt{slurm.conf}) & Compute Nodes & \texttt{slurmd} & SlurmdUser (normally root user) & At job termination \\
    \midrule
    EpilogSlurmctld (specified in \texttt{slurm.conf}) & Head Nodes or Master Nodes (where \texttt{slurmctld} runs) & \texttt{slurmctld} & SlurmctldUser & At job termination \\
    \bottomrule
    \end{tabular}
\end{table*}


In the first demonstration of our deployment on the \gls{hpc} system at \gls{lrz}, we integrated the plugins with the SLURM controller and daemon using the following concrete steps:
\begin{itemize}
    \item \glspl{qpu} are defined and configured as \glspl{gres} (\texttt{--gres}) in SLURM. For example, we put the specification in \texttt{slurm.conf} and \texttt{gres.conf}:
    
\begin{lstlisting}[language=bash, label={lst:slurm-gres}]
# In slurm.conf
GresTypes=gpu,qpu
NodeName=node[01-N] Gres=gpu:1,qpu:1 ...
    
# In gres.conf
Name=gpu File=/dev/gpudev0
Name=qpu Type=general Count=1
\end{lstlisting}

    Depending on the number of \glspl{qpu} and how we allocate them, \texttt{--gres} can be set to specify a different number of available shared resources. This configuration enables SLURM to control \gls{qpu} allocation and identify which nodes support \gls{qpu}-required jobs.

    \item In the background, \texttt{slurmctld} and \texttt{slurmd} operate as event-based models, in which jobs are queued, allocated resources, executed, and terminated in different phases. Thereby, to trigger the connection and offload quantum tasks to the quantum server, we leverage this mechanism to facilitate quantum task offloading from the \gls{hpc} side. SLURM's workflow can be intercepted at the predefined phases shown in Table~\ref{tab:slurm-phases}.
\end{itemize}

Our plugins leverage the phases of job allocation and termination to enable and disable quantum task offloading. For instance, we set the plugin triggered by \texttt{PrologFlags=Alloc}. At this stage, the quantum daemon (\texttt{qdaemon}) is initialized, the \gls{qpu} is set to ``occupied'', and one of the components in \gls{mqss}, the \textit{MQSS Client}, establishes the connection between \gls{hpc} and 
the rest of the stack via \texttt{qdaemon}. This authenticates the access to the corresponding/requested quantum devices.

At the termination phase, upon job completion or cancellation, the plugin is triggered to stop the connection. This stage ensures a ``clean'' exit by stopping the \texttt{qdaemon}, the connection between two sides, and releasing the allocated quantum resources for the next user.

Instead of simply implementing this \gls{hpcqc} job submission concept using Prolog/Epilog, we can use SPANK to provide more granular control in C. Prolog/Epilog is effective for high-level orchestration, while SPANK can be developed as shared libraries in C, allowing for in-depth environment customization. The \textit{MQSS SLURM Plugin Suite} also demonstrates SPANK implementation to be a more transparent interaction with SLURM job steps. Our dual approach ensures that the suite remains both lightweight and capable of handling the unique demands of quantum-classical hybrid workloads.

\section{Deployment on SuperMUC-NG with IQM Q-Exa} 

\begin{figure*}[ht]
  \centering
  \includegraphics[width=\textwidth]{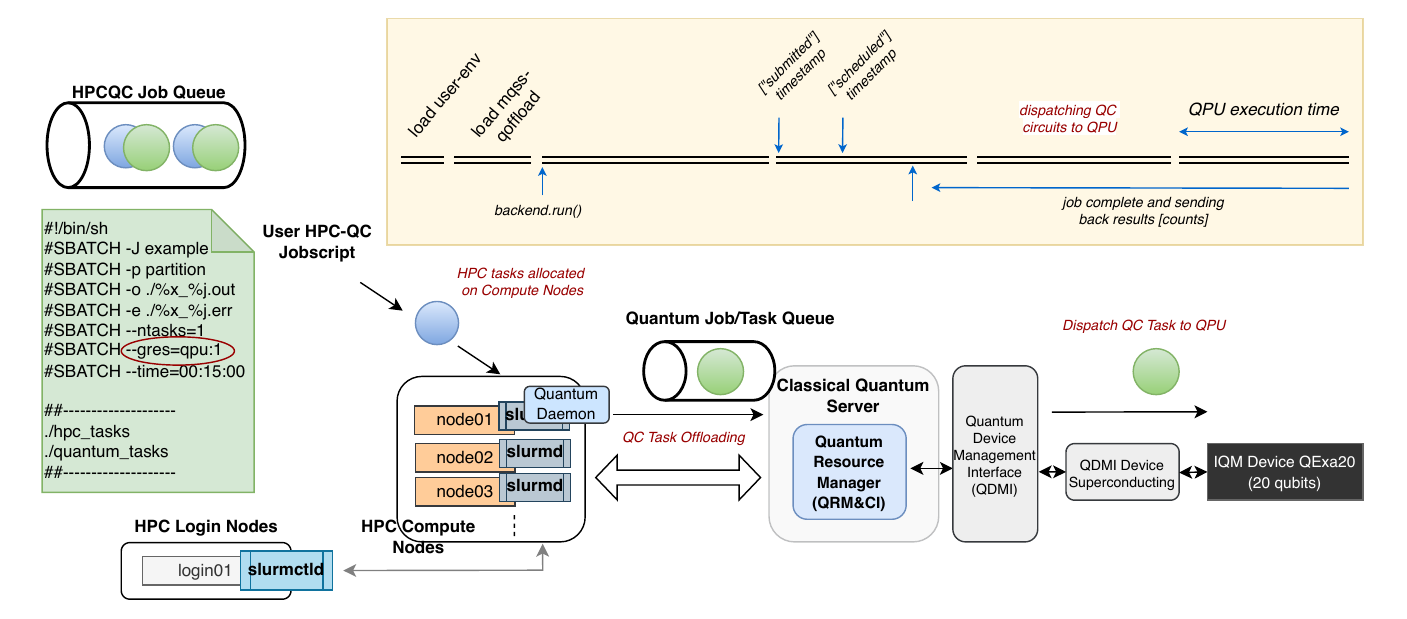}
  \caption{An illustration of a hybrid HPCQC job and the working pipeline between HPC and QC with MQSS.}
  \label{fig:hpcqc-exe-pipeline}
\end{figure*}

We deployed \gls{mqss} on an \gls{lrz} testbed system (named BEAST) and in production on SuperMUC-NG, 
integrating \gls{iqm}'s 20-qubit superconducting processor, Q-Exa, as an external accelerator. 
The \gls{qpu}, along with its dedicated control stack and secure operational domain, is provided and maintained by \gls{iqm}, ensuring reliable operation and compliance with vendor-specific requirements. 
\gls{mqss} operates across multiple network and security zones to maintain both performance and security: \textit{MQSS Clients} run on the allocated compute nodes within the \gls{hpc} environment, \gls{qrm} executes in a secured orchestration zone, and \gls{qdmi} communicates with the vendor's control stack to obtain real-time device status, including topology, fidelity metrics, calibration windows, and availability. 

From the end-user perspective, hybrid jobs are submitted as a single SLURM job that seamlessly combines classical and quantum computation. 
Classical pre-processing and post-processing tasks are executed on SuperMUC-NG, while quantum kernels are offloaded to Q-Exa through \gls{mqss}. 
The stack handles just-in-time compilation, device-aware optimization, and kernel dispatch transparently, so users interact with the system much like they would with conventional accelerators such as \glspl{gpu}. 
In operational use, we observed that the combined workflow behaves predictably: the SLURM controller allocating resources simultaneously covers both the requested compute nodes and the \gls{qpu}, while \gls{mqss} ensures that quantum execution proceeds without stalling or manual intervention. 
This integration demonstrates the feasibility of treating a remote superconducting quantum processor as a scheduler-managed accelerator, enabling researchers to run complex hybrid \gls{hpcqc} workflows in a production environment with minimal overhead and maximal usability.

Figure~\ref{fig:hpcqc-exe-pipeline} shows an illustration of HPCQC job submission on our system. The illustration highlights a jobscript example and the pipeline of SLURM plugins on HPC nodes that make a connection to the quantum resource manager and related components to reach the requested QPU. Given a hybrid application, users can explicitly specify HPC and quantum tasks in the job script. Alternatively, users can implement both tasks in a single code. With the SLURM utility commands, we can submit the job, e.g., via \texttt{sbatch} or \texttt{salloc} for interactive jobs.

\begin{itemize}
    \item When a job is allocated and starts, our SLURM plugin is triggered. It checks if the job requires quantum resources, then initializes the quantum daemons to establish the \gls{qdmi} connection.
    \item At the execution phase, classical tasks are run on the HPC and quantum tasks are offloaded to the quantum server. After processing the task at \gls{qrm}, it dispatches the quantum circuits for execution on the \gls{qpu} via \gls{qdmi}. The interaction between the two sides can be tightly integrated. Once the quantum computation has finished, the results are sent back to the application via the same route. Since \gls{qpu} is not a native resource and it is shared among users, this approach is practical for a hybrid \gls{hpcqc} workload.
    \item After the job is complete, the SLURM plugin ensures to stop the connection, quantum daemon, and we can release the \gls{qpu} resource.
\end{itemize}

There are two queues of jobs, as we can see in Figure~\ref{fig:hpcqc-exe-pipeline}, one on \gls{hpc} and one on \gls{qc}. In the current deployment, jobs are received from the \gls{hpc} side, where the classical resource is prioritized. Then, \gls{qpu} is checked to let SLURM decide job allocation. This approach comes with shortcomings, such as long blocking of a resource slot on one or both sides. For example, \gls{hpc} waits for \gls{qc} execution or vice versa. Therefore, this opens challenges for studying coordinated scheduling between two-sided schedulers or multilevel scheduling algorithms. In the following section, we present experiments and results from our practical deployment as lessons learned. We present the hybrid execution pipeline as highlighted in the yellow block that we also can see in Figure~\ref{fig:hpcqc-exe-pipeline}. The pipeline shown details the steps involved when a quantum task or quantum circuit is submitted from the HPC. This pipeline will align with the experimental results presented in the next section as we evaluate practical hybrid applications. We focus on evaluating the execution time between the HPC and QC, as well as the latency times between steps when quantum circuits are offloaded to the quantum server. The quantum server processes the circuits, and then the QDMI, before the circuit is run on the QPU. In general, these steps are characterized above as:
\begin{itemize}
    \item \texttt{load user-env}: Load the necessary libraries and environment.
    \item \texttt{load mqss-qoffload}: Load components of the software stack on the compute node to make a connection with the QC via MQSS to enable offloading quantum tasks.
    \item \texttt{["submitted" timestamp]}: Indicates the timestamp when the circuit on the Quantum side is submitted.
    \item \texttt{["scheduled" timestamp]}: Indicates the timestamp when the circuit is scheduled to QPU.
    \item \texttt{QPU execution time}: Denotes the runtime on QPU, which is roughly measured when the circuit is dispatched on QPU and completes to give back results. In which we can check by \texttt{["complete" timestamp]}.
\end{itemize}

\section{Performance and operational observations} 

In this section, we present experiments on two hybrid benchmarks, VQE~\cite{cerezo2021vqe} and QAOA~\cite{fakhimi2023qaoa}. VQE estimates the expectation value of a Hamiltonian, requiring multiple quantum circuit runs per iteration to measure different terms and bases, in addition to classical pre-processing steps such as Hamiltonian grouping and initial-state preparation (e.g., Hartree-Fock~\cite{google2020hartreefock}). In contrast, the standard QAOA ansatz for MaxCut involves a diagonal Hamiltonian, allowing each iteration to be executed with a single quantum circuit and with minimal classical pre-processing. We employ a vanilla ansatz composed of interleaved, parameterized cost and mixing layers. The algorithm is applied to the MaxCut problem, a classical graph optimization task that partitions a graph into two sets to maximize the number of edges connecting them. A key hyperparameter of QAOA is the number of layers, which is the number of QAOA layers, each consisting of a cost and a mixing unitary with independent parameters. We select this depth dynamically by running the algorithm with fewer optimization iterations across different layer counts, revealing the trade-off between under-parameterization and circuit depth. For this use case, the circuits are configured to solve a four-node MaxCut instance using two QAOA layers. The key differences of the two approaches affect both the frequency of quantum runs and the computational requirements of classical computation, supporting the understanding of performance bottlenecks across hardware and software components.

\begin{figure}[t]
  \centering
  \includegraphics[width=\columnwidth]{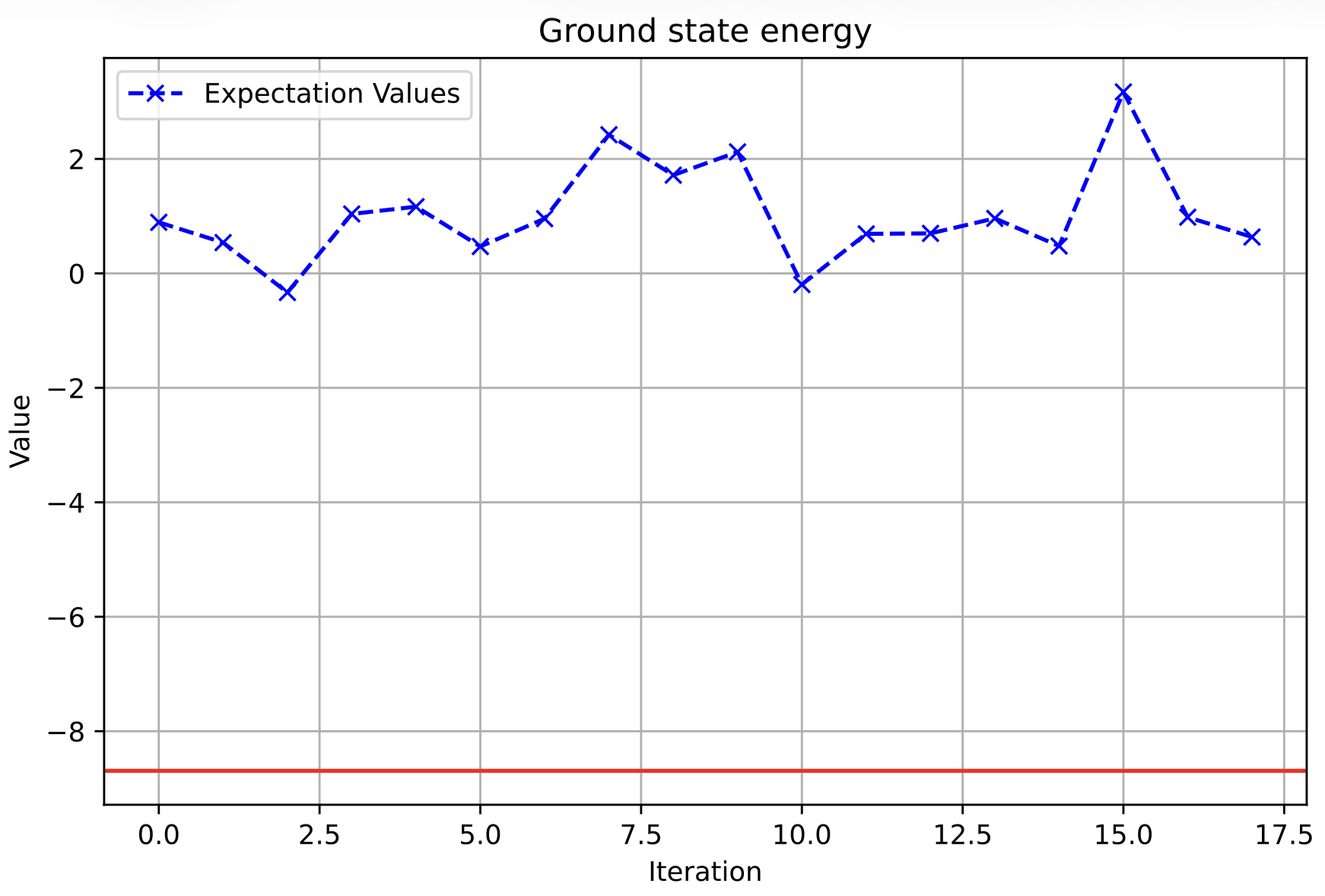}
  \caption{The energy-convergence results for the VQE benchmark performed on the Q-Exa 20-qubit device.}
  \label{fig:vqe-perf-result}
\end{figure}

\begin{figure}[t]
  \centering
  \includegraphics[width=\columnwidth]{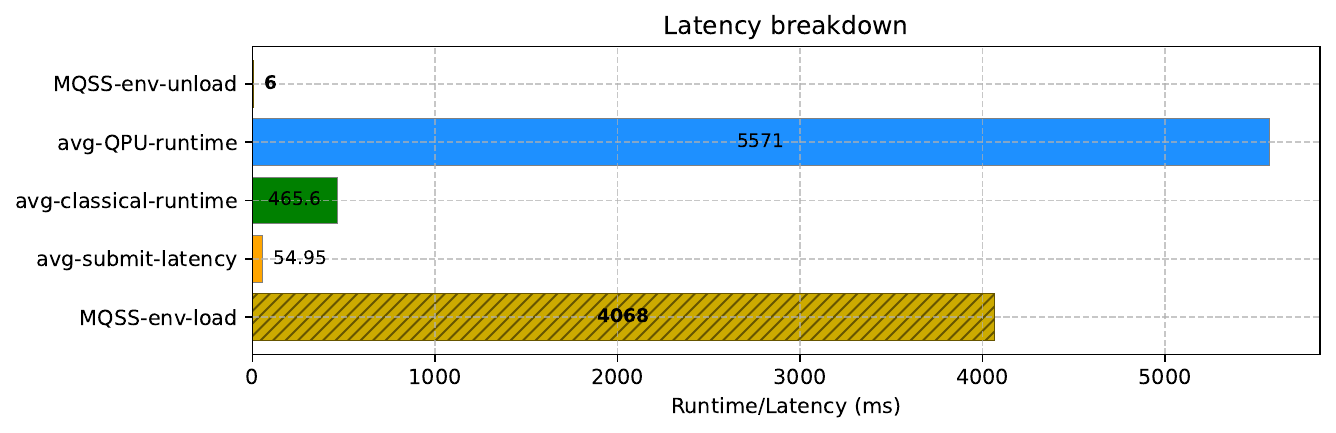}
  \caption{Latency of the hybrid VQE-execution pipeline on the HPCQC integrated system at LRZ.}
  \label{fig:vqe-latency-result}
\end{figure}

\begin{figure*}[t]
  \centering
  \includegraphics[width=\textwidth]{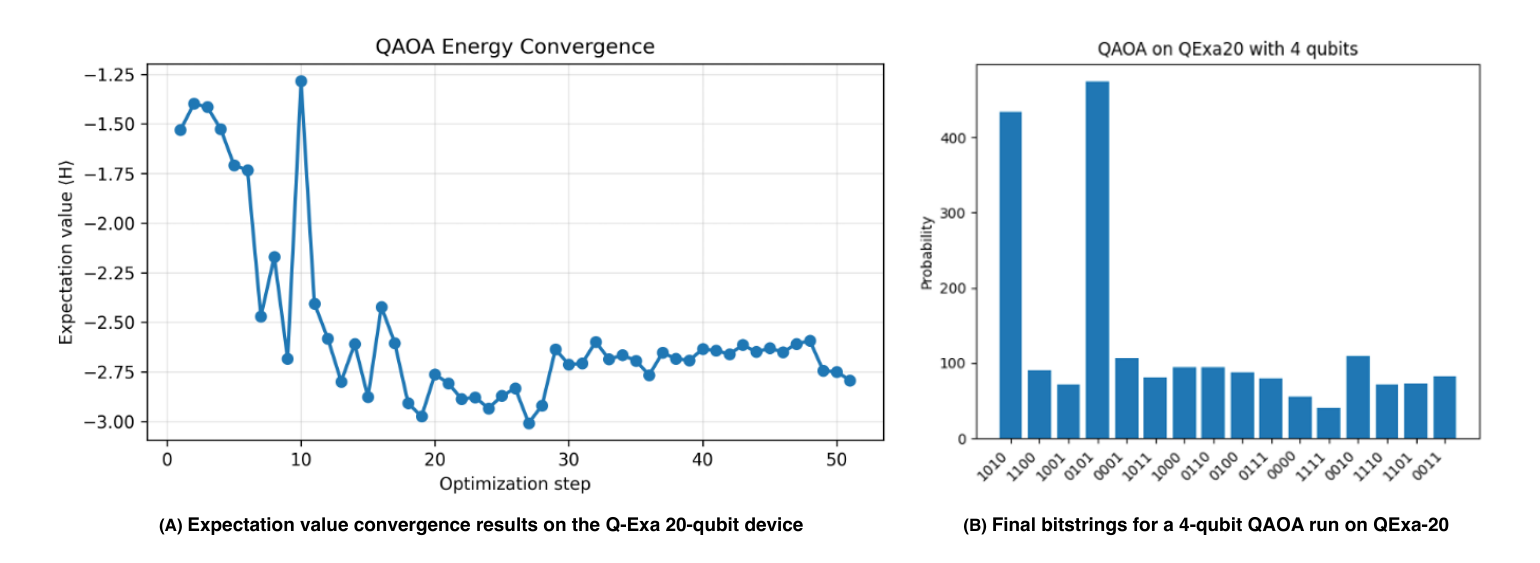}
  \caption{The experimental results of QAOA showing (A) expectation value of convergence on Q-Exa and (B) the final bitstrings for the given 4-qubit QAOA run.}
  \label{fig:qaoa-perf-result}
\end{figure*}

\begin{figure}[t]
  \centering
  \includegraphics[width=\columnwidth]{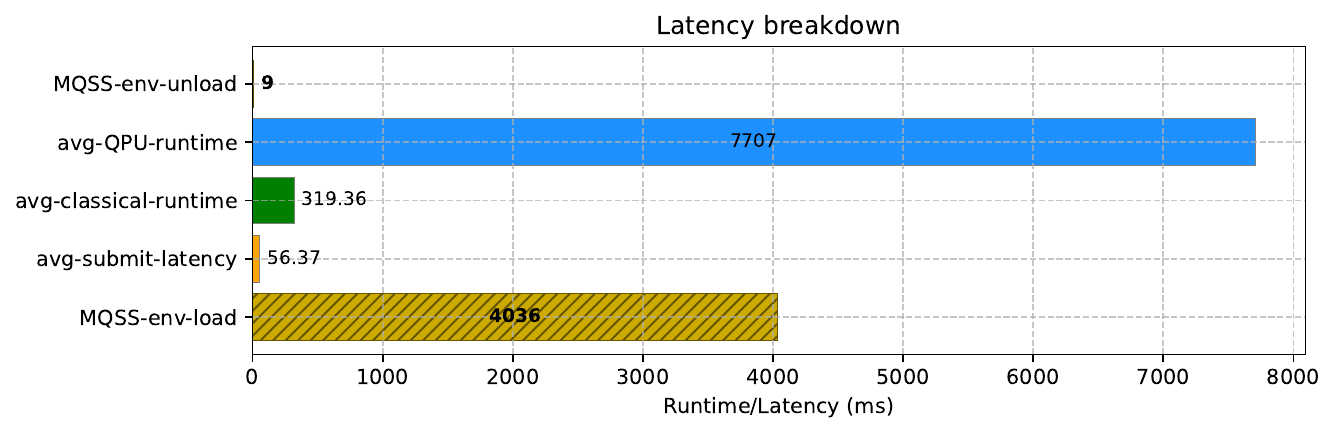}
  \caption{Latency of the hybrid QAOA-execution pipeline on the HPCQC integrated system at LRZ.}
  \label{fig:qaoa-latency-result}
\end{figure}

\subsection{VQE Benchmark}
As mentioned, to ease benchmarking and profiling execution time, we use a precomputed Hamiltonian for the $H2$ molecule at a bond length of $0.7414 Angstroms$. This molecular Hamiltonian is then decomposed as a linear combination of initial Pauli terms and their coefficients. 
Following the Jordan-Wigner mapping, the VQE circuit size is set to $4$ qubits, after the qubit encoding. However, the qubit scaling could also differ for different choice of bases and qubit mappings. 

Figure~\ref{fig:vqe-perf-result} shows the results on the QExa $20$-qubit device. We run the VQE benchmark with $17$ iterations. Based on the initial configuration, each iteration performs $16$ circuits corresponding to the $14$ Pauli terms in the problem design. For instance, the initialized Pauli terms and offsets are given as follows.

\begin{lstlisting}[language=Python, label={lst:vqe-hamiltonian}]
H = [
        ("ZIII", 0.17110545123720233),
        ("IZII", 0.17110545123720225),
        ("ZZII", 0.16859349595532533),
        ("YXXY", 0.04533062254573469),
        ("YYXX", -0.04533062254573469),
        ("XXYY", -0.04533062254573469),
        ("XYYX", 0.04533062254573469),
        ("IIZI", -0.22250914236600539),
        ("ZIZI", 0.12051027989546245),
        ("IIIZ", -0.22250914236600539),
        ("ZIIZ", 0.16584090244119712),
        ("IZIZ", 0.16584090244119712),
        ("IIZZ", 0.12051027989546245),
        ("ZZZZ", 0.1743207725924201),
    ]

\end{lstlisting}

The results in Figure~\ref{fig:vqe-perf-result} show only the execution of $17$ iterations, in which this does not reach an acceptable energy convergence for this VQE task because this use case requires hundreds of iterations to reach a better result that makes sense for the H2 simulation problem. However, here we are interested in the workflow and how latency breaks down into different steps. The total runtime of 17 iterations is around 35.67 minutes, in which the VQE benchmark's energy can only reach -0.3345 as minimum.

This setup highlights that each iteration of the VQE workflow will execute 16 corresponding circuits as the submitted quantum tasks to QPU. Figure~\ref{fig:vqe-latency-result} shows the latency of our hybrid execution pipeline, which is divided into five key intervals corresponding to the hybrid execution pipeline as follows.

\begin{itemize}
    \item \texttt{MQSS-env-load}: the time to load the components of the software stack to enable quantum task offloading on the HPC side. It is required only once, when the SLURM job is launched.
    \item \texttt{avg-submit-latency}: indicates the average time when a circuit is offloaded from the HPC side until it is queued and scheduled on the QC side.
    \item \texttt{avg-classical-runtime}: indicates the average execution time on the classical side of each iteration.
    \item \texttt{avg-QPU-runtime}: indicates the average execution time of each quantum circuit run on the real QPU.
    \item \texttt{MQSS-env-unload}: highlights the unloading time that makes sure cleaning up the environment, stops the connection, and releases the QPU resource.
\end{itemize}

In Figure~\ref{fig:vqe-latency-result}, we can see that the submit latency is negligible compared to the average execution time of a circuit on QPU as well as the classical runtime. In principle, the execution time of a circuit or classical computation depends on the nature of the problem and the number of computations required. However, the experimentally measured values here still objectively reflect our HPCQC integration pipeline in practice. Simultaneously, compared to large computations or real-world problems, the latency value here is still much smaller than the execution time required on QPU, CPU, or GPU. As a result in average, the \texttt{avg-submit-latency} is 54.95ms compared to 465.6ms of \texttt{avg-classical-runtime} and 4068ms of \texttt{avg-QPU-runtime}.

\subsection {QAOA Benchmark}

We have designed the QAOA benchmark to target the MaxCut problem. A small graph example is, with $4$ nodes and $4$ edges, to achieve reasonable accuracy. The benchmark uses a vanilla QAOA ansatz, composed of an initial-state preparation (a Hadamard layer for all qubits to prepare an equal superposition) interleaved with \textit{mixing and cost}~\cite{fakhimi2023qaoa} unitaries. The mixing and cost unitaries are also repeated $L$ times, each with different parameters, where $L$ indicates the number of QAOA layers. 

Figure~\ref{fig:qaoa-perf-result} shows the analysis of the expectation value calculation during QAOA optimization. In which its sub-figure (B) highlights the bit-strings sampled in the final iteration, with a clear peak for the two bit-strings ($0101$, $1010$) that are the optimal solution for the MaxCut problem for the selected graph.

Similar to the latency breakdown of the VQE benchmark, Figure~\ref{fig:qaoa-latency-result} shows the average runtime and latency for a single circuit execution in an iteration of the QAOA benchmark. In practice, we run the entire execution for about $50$ iterations to achieve convergence. To show latency, we calculate the average runtime of circuits by measuring the time between dispatching each circuit to the QPU and receiving its probability result from the quantum server. This preliminary result highlights the execution time, including latency effects, in our quantum software stack. As we can see, the submit latency and time to load/unload the quantum software stack environment remain roughly the same (56.37ms for the \texttt{avg-submit-latency}, 4036ms for \texttt{MQSS-env-load}, and 9ms for \texttt{MQSS-env-unload}). Loading/unloading the MQSS software environment only needs to be done once. The QPU runtime for executing the circuits depends on how the quantum circuits are submitted.

\section{Limitations and Discussion}
The paper's results demonstrate that hybrid \gls{hpcqc} workflows can be executed on a production system with negligible latency. Nevertheless, our current deployment is an early integration milestone, and several aspects of its limitations warrant discussion.
\paragraph{Hybrid workflow coverage}
We used VQE and QAOA as representative workloads to evaluate the hybrid execution pipeline. The experiments highlight exchange data between QPU and HPC once per iteration over many short-lived circuits. The results show that the overhead is negligible compared to the total execution time; however, this evaluation does not cover all possible hybrid workflows, and other classes of applications may exhibit different performance characteristics. We therefore expect that investigating the overheads will show a conservative upper bound, and that a systematic study across the full spectrum of coupling granularities remains future work.
\paragraph{QPU availability, sharing, and reconfiguration}
The way a \gls{qpu} is shared follows directly from \gls{gres} configuration. In our demonstrated configuration, the example shows that only a single \gls{qpu} is available. It is exposed to SLURM as one \gls{gres} unit (\texttt{qpu:1}) and is effectively shared by all users, with SLURM serializing access to the device across hybrid jobs. When multiple \glspl{qpu} are available, the \gls{gres} can be configured with the corresponding count (e.g., \texttt{--gres=qpu:2,3,4}), allowing SLURM to allocate distinct devices to concurrent jobs. On the quantum devices side, the quantum jobs offloaded from the allocated \gls{hpc} nodes are delegated to the \gls{qrm} queue, which handles their queuing and dispatch to the devices. \emph{Quantum availability} can be derived from periodic \gls{qdmi} device-status queries, which expose topology, gate fidelities, and calibration windows. Device reconfiguration and recalibration are currently treated by the vendor control stack. Finer-grained, concurrent sharing of a single \gls{qpu} across multiple SLURM jobs is an open item for future deployments.
\paragraph{Coordination between the two schedulers}
Our current HPCQC model prioritizes the classical allocation: jobs are admitted from the \gls{hpc} side, and the \gls{qpu} is then checked before dispatch. This keeps the integration portable, but it does not yet jointly optimize the use of classical and quantum resources. As a result, under high load, a reserved classical allocation may idle while it waits on the \gls{qpu}, or vice versa. Besides, a more detailed characterization of QPU utilization at scale, beyond the per-circuit latency breakdown reported in this paper, is part of our ongoing work.
\paragraph{Generality of latency result}
For the generality of the latency result, two-level schedulers depend on per-task overhead. This would make submit-and-dispatch latency non-negligible. Our current architecture exposes QPUs as \gls{gres} and delegating device-level sequencing to a second-level scheduler remains valid, but the quantitative latency results should be read in light of this dependency. Further work is needed to systematically study the overheads across a broader spectrum of coupling granularities.

\section{Conclusions} 
Our work demonstrates a practical and portable approach to integrating quantum processors into SLURM-managed \gls{hpc} centers without invasive scheduler changes. 
The \gls{mqss} architecture, including \gls{qrm}, \gls{qdmi}, and the \textit{MQSS SLURM Plugin Suite}, allows \gls{hpc} centers to offer quantum accelerators with the same operational model and user experience as \glspl{gpu} or \glspl{fpga}. 
We credit \gls{iqm} for the Q-Exa hardware and \gls{mqv} along \gls{lrz} and \gls{tum} for the development of \gls{mqss}. 
The software is designed to be portable to Blue Lion (\gls{lrz}'s forthcoming \gls{hpe} Cray system) and other SLURM-managed installations, enabling CUG centers to adopt hybrid \gls{hpcqc} workflows with minimal disruption to existing user practices. This work provides a blueprint for scheduler-level integration of emerging quantum resources into mainstream \gls{hpc} environments.

We emphasize that hybrid execution on the integrated SuperMUC-NG + Q-Exa platform introduces no significant latency overhead in a co-located \gls{hpcqc} setup. \gls{qrm}'s just-in-time compilation, guided by \textit{QDMI Device} metrics, is sufficiently fast and may be overlapped with other job stages so that the quantum operations do not introduce observable idle time at the scheduler level. 
The SLURM plugin suite ensures that the hybrid jobs begin only when the \gls{qpu} is ready. 
Daily operational needs, such as recalibration, are integrated into the availability model to minimize disruptions.


\begin{acks}
This work is supported by the German Federal Ministry of Research, Technology and Space (BMFTR) with the grants 13N16063 (Q-Exa), 13N16690 (Euro-Q-Exa), 101114305 (Millenion), 10111394-6 (OpenSuperQPlus), and the Bavarian State Ministry of Science and the Arts (StMWK) through funding, as part of MQV, Q-DESSI.

The authors would also like to acknowledge the use of DeepL and ChatGPT for grammar checking and improving the clarity of language.
\end{acks}

\bibliographystyle{ACM-Reference-Format}
\bibliography{references}

\end{document}